\documentclass{PoS}

\title{Lower dimensional Yang-Mills theory \\ as a laboratory to study the infrared regime}

\ShortTitle{Lower dimensional Yang-Mills theory as a laboratory to study the infrared regime}

\author{Reinhard Alkofer \\
        Institut f\"ur Physik,
Karl-Franzens-Universit\"at,
Universit\"atsplatz 5,
A-8010 Graz, Austria}

\author{Christian S. Fischer \\
         Institut f\"ur Physik, TU Darmstadt,
        Schlossgartenstr. 9,
        D-64289 Darmstadt, Germany}

\author{Markus Q. Huber \\
        Institut f\"ur Physik,
Karl-Franzens-Universit\"at,
Universit\"atsplatz 5,
A-8010 Graz, Austria}

\author{\speaker{Kai Schwenzer} \\
        Institut f\"ur Physik,
Karl-Franzens-Universit\"at,
Universit\"atsplatz 5,
A-8010 Graz, Austria\\
        E-mail: \email{kai.schwenzer@uni-graz.at}}

\abstract{Lattice studies of the infrared regime of gauge theories are
complicated by the required extensive limits, the performed gauge fixing and
the demand for high statistics. Using a general power counting scheme for the
infrared limit of Landau gauge SU(N) Yang-Mills theory in arbitrary dimensions
we show that the infrared behavior of Greens functions is both qualitatively
and quantitatively similar in two, three and four spacetime dimensions.
Therefore, lower dimensional lattice simulations can serve as a simplified
laboratory to analyze the presently applied approximations and to obtain first
results for higher correlation functions.}

\FullConference{The XXV International Symposium on Lattice Field Theory\\
		 July 30 - August 4 2007\\
		 Regensburg, Germany}

\begin{document}

\paragraph*{Introduction:}

Lattice studies of the infrared (IR) regime of Yang-Mills theory, see e.g.
\cite{Sternbeck:2005tk}, are important to understand its genuinely
non-pertubative aspects. This is particularly promising since the detailed
mechanism for gluon confinement could be connected to directly measurable
correlation functions of the colored fields within the Kugo-Ojima
\cite{Kugo:1979gm}  and Gribov-Zwanziger \cite{Gribov:1977wm} scenarios. These 
rely on a strong IR increase of the ghost dressing function that was predicted
by continuum methods
\cite{vonSmekal:1997is,Zwanziger:2001kw,Lerche:2002ep,Alkofer:2000wg} and has
been clearly confirmed in lattice simulations \cite{Sternbeck:2005tk}. 
Combined efforts via all available methods such as Dyson-Schwinger equations
(DSE) \cite{vonSmekal:1997is,Zwanziger:2001kw,Lerche:2002ep,Alkofer:2000wg},
renormalization group  (RG) techniques \cite{Pawlowski:2003hq} and lattice
gauge theory studies \cite{Sternbeck:2005tk} led during the last years to a
coherent picture of the IR regime in Landau gauge. The continuum
approaches allow in particular to analyze the IR scaling limit of general
Greens functions \cite{Alkofer:2004it}. These are compatible with a dominance
of the gauge fixing part of the action, as argued for in
\cite{Zwanziger:2003cf}. A simultaneous analysis of DSE and RG methods allowed
furthermore to show  that this IR fixpoint is unique \cite{Fischer:2006vf}. \\
Recently, challenging lattice results by several groups on larger lattice sizes
\cite{Bowman:2007du} suggested that the IR scaling of the gluon propagator
might be weaker than predicted by continuum methods and does not show the IR
vanishing required within the Gribov-Zwanziger scenario. The finite size
corrections have been previously estimated within a Dyson-Schwinger study on a
compact manifold \cite{Fischer:2005ui} and suggest that despite significant
corrections the expected IR behavior should be visible with the quite
remarkable lattice sizes employed. However, the four dimensional studies still
feature significant statistical fluctuations in the IR which are less
pronounced in two \cite{Maas:2007uv} and three dimensions
\cite{Cucchieri:2006tf} where deviations from the continuum predictions are not
observed. \\
The discrepancy of the new lattice data with the continuum results can
basically have two reasons. First, it is possible that the IR regime requires
even more refined numerical methods on the lattice side. As we will argue
below, in this case lower dimensional studies should allow to test the involved
approximations in a substantially simpler setting. This conclusion stems from a
study of the dependence of the IR limit on the spacetime dimension $d$.  We
establish  a manifest power counting scheme for general vertex functions of
Landau gauge  Yang-Mills theory in arbitrary dimensions and find qualitatively
similar results  as in the four-dimensional analysis given in
\cite{Alkofer:2004it}. \\
The other reason for  the discrepancy could be that the IR behavior predicted
by continuum methods - and in particular the value of the IR scaling
parameter $\kappa$ - suffers from the presently studied approximations. In an
approximation based on the propagator DSEs and using a bare ghost-gluon vertex
it takes a value $\kappa\approx0.595$ in four dimensions. A possible dressing
of this vertex could change this value \cite{Lerche:2002ep}, and the continuum
prediction would be compatible with the recent lattice results if the IR
exponent $\kappa$ is slightly smaller. Therefore, we also discuss the
dependence of the dynamical building blocks in the Dyson-Schwinger equations on
$\kappa$.  

\paragraph*{IR exponents for arbitrary $d$:}

\begin{figure}[h]
\centerline{\includegraphics[width=\textwidth]{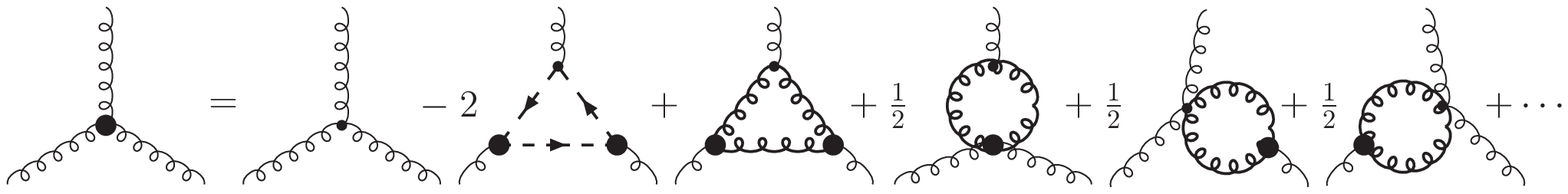}}
\caption{\label{fig:3g-DSE-skelexp-2} 1-loop part of the skeleton expansion of the DSE for the three-gluon vertex.}
\end{figure}

\begin{figure}
\centerline{\includegraphics[width=0.8\textwidth]{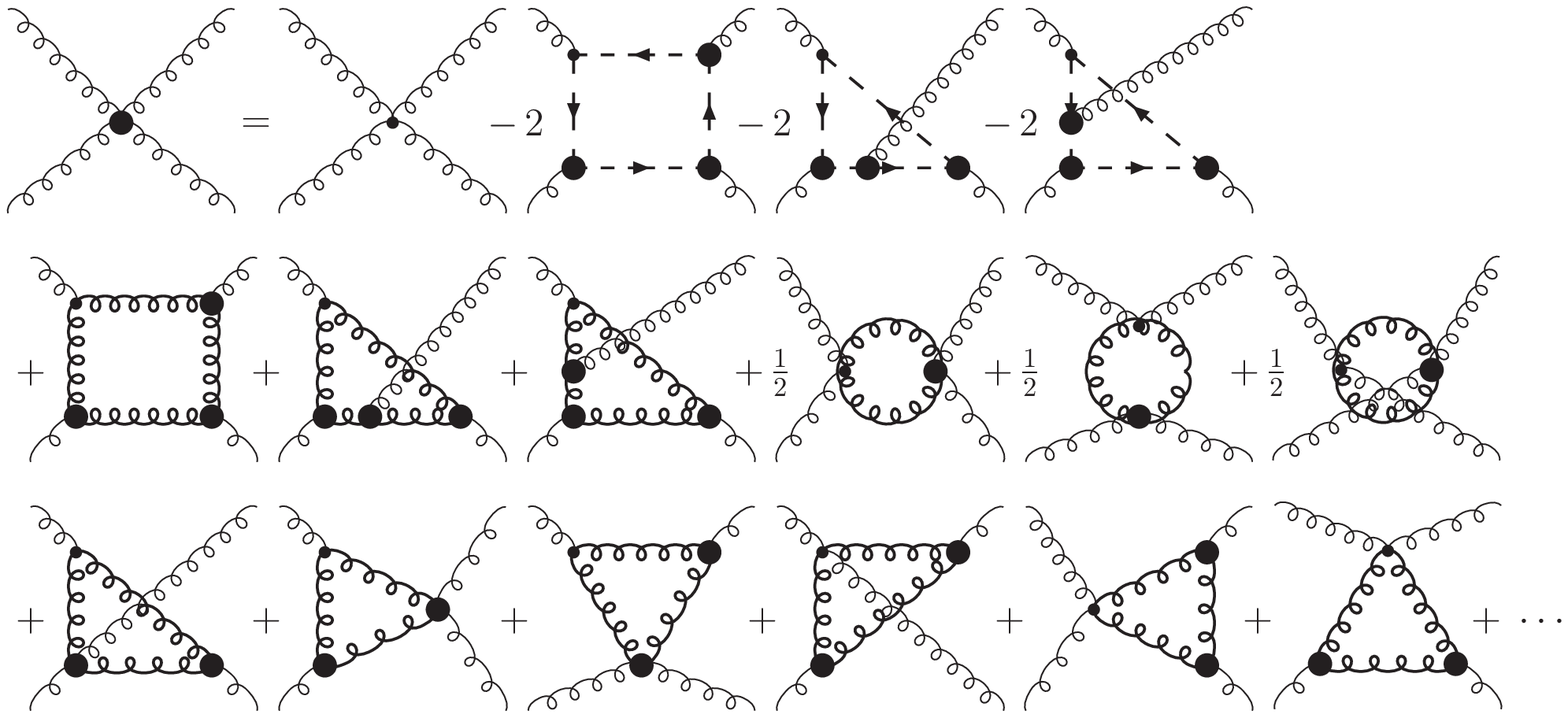}}
\caption{\label{fig:4g-DSE-skelexp} 1-loop part of the skeleton expansion 
of the DSE for the four-gluon vertex.}
\end{figure}

In the following we will perform a scaling analysis for the IR regime of
Yang-Mills  theory in arbitrary dimensions. In contrast to the generic IR limit
we study here the  limit where the coupling $g$ is kept fixed and has no
inherent scaling dependence, which is more directly accessible in lattice
simulations and has been analyzed  in \cite{Maas:2007uv,Cucchieri:2006tf}. As
in four dimensions \cite{Alkofer:2004it}, the starting point for the IR
analysis is the non-renormalization of the ghost-gluon vertex in Landau gauge
\cite{Taylor:1971ff},  which implies a finite vertex in the IR
\cite{Alkofer:2004it}.  This property depends purely on the transversality of
the gluon propagator and is therefore valid in arbitrary dimensions.
\begin{figure}
\centerline{\includegraphics[width=0.7\textwidth]{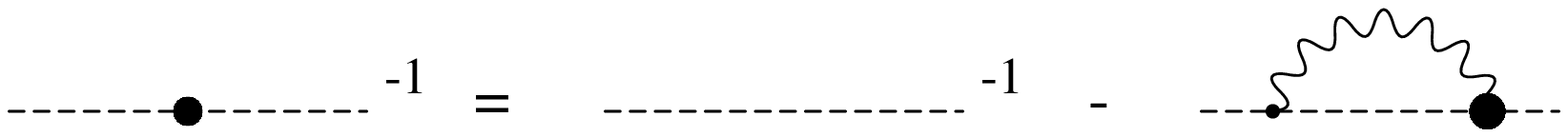}}
\caption{\label{fig:gh-DSE-again} The DSE for the ghost propagator.}
\end{figure}
The IR behavior of the propagators and vertices well below its inherent
scale $g^{2/(4-d)}$ (respectively $\Lambda_{QCD}$ in $d=4$) is determined via
renormalization group arguments by scaling relations. The propagators  of the
gluons and ghosts 
\begin{equation}
D_{\mu\nu}(p^2)=\left( \delta_{\mu \nu}-\frac{p_\mu p_\nu}{p^2} \right) 
\frac{Z(p^2)}{p^2} \; , \qquad D^G(p^2)=-\frac{G(p^2)}{p^2} \, ,
\end{equation}
are given in terms of dressing functions whose IR behavior is described 
by a power law ansatz
\begin{equation}\label{eq:power-laws-props}
Z(p^2)=c_{0,2} \cdot(p^2)^{\delta_{0,2}} \; , \quad G(p^2)= c_{2,0} \cdot(p^2)^{\delta_{2,0}} \, ,
\end{equation}
and similar for the vertices. Here we denote the IR exponent of  a vertex with
$2n$ ghost and $m$ gluon legs by $\delta_{2n,m}$ and the corresponding 
coefficient by $c_{2n,m}$. Whereas this coefficient is a constant for the
propagators  it is generally a function of $2n\!+\!m\!-\!1$ momentum ratios. \\
For the integral on the right hand side of the ghost propagator Dyson-Schwinger
equation,  cf. fig. \ref{fig:gh-DSE-again}, one can use the standard
expression  \cite{Lerche:2002ep,Anastasiou:1999ui} 
\begin{equation}
\label{eq:2-point-integral}
\int \frac{d^dq}{(2\pi)^d} (q^2)^{\nu_1}((q-p)^2)^{\nu_2}=(4\pi)^{-\frac{d}{2}}
	\frac{\Gamma(\frac{d}{2}+\nu_1)\Gamma(\frac{d}{2}+\nu_2)\Gamma(-\nu_1-\nu_2-\frac{d}{2})}
		{\Gamma(-\nu_1)\Gamma(-\nu_2)\Gamma(d+\nu_1+\nu_2)} (p^2)^{\frac{d}{2}+\nu_1+\nu_2}
\end{equation}
which shows that it scales proportional to the external momentum. The left hand
side of the ghost DSE, which consists only of the inverse dressed ghost 
propagator, is proportional to $(p^2)^{-\delta_{2,0}+1}$, where the $1$ is the
canonical  dimension. The scaling dimensions on the right hand side are  $d/2$
from the integral, $\delta_{0,2}-1$ from the gluon propagator,
$\delta_{2,0}-1$  from the ghost propagator, $1/2$ from the bare ghost-gluon
vertex, and $1/2$  from the dressed ghost-gluon vertex which features no
anomalous scaling.  Defining the parameter $\kappa$ as $\kappa :=
-\delta_{2,0}$, this condition on the IR exponents yields  in four dimensions
the well-known result $\delta_{0,2}=2 \kappa$. In $d$ dimensions we have
\cite{Zwanziger:2001kw,Lerche:2002ep}
\begin{equation}
  \delta_{2,0}=-\kappa \qquad , \qquad \delta_{0,2}=2\kappa+2-\frac{d}{2} \, .
\end{equation}
In order to transform the infinite hierarchy of Dyson-Schwinger equations into
a closed  system we perform a skeleton expansion. This yields an infinite tower
of graphs involving  only primitively divergent vertices which we will analyze
in a first step. The first order  of the skeleton expansion of the DSE for the
three-gluon vertex is depicted in fig.  \ref{fig:3g-DSE-skelexp-2}.
 \\
We start with the ghost triangle which turns out to be one of the IR
leading diagrams. Subtracting the canonical dimension $1/2$ we get from the IR
counting rules the anomalous IR exponent:
\begin{equation}
\label{eq:3g}
\delta_{0,3}^{gh\Delta}=\frac{d}{2}+3(-\kappa-1)+3\,\frac{1}{2}-\frac{1}{2}
=-3\kappa+\frac{d}{2}-2 \, .
\end{equation}
For the four-gluon vertex we can apply the same procedure. In fig.
\ref{fig:4g-DSE-skelexp} we  show the first order of its skeleton expansion.
Here, again the ghost rectangle gives one of the IR  dominant contributions:
\begin{equation}
\delta_{0,4}^{gh\square}=\frac{d}{2}+4(-\kappa -1) +4 \,\frac{1}{2}
=-4\kappa +\frac{d}{2}-2 \, .
\end{equation}
With these results for the scaling of the primitively divergent vertices it is
easy to check that the other diagrams in the DSEs fig.
\ref{fig:3g-DSE-skelexp-2} and \ref{fig:4g-DSE-skelexp} have either the same
scaling or are subleading in the IR. Similarly one can analyze the IR exponent
of an arbitrary $n$-point function. This is done by counting the IR exponents
of the individual building blocks that arise in a general loop correction. The
resulting expressions can be simplified via topological identities. It is then
straightforward to show that the leading IR exponent of an $n$-point function
is given by the simple expression \cite{Huber:2007kc}
\begin{equation}
\label{eq:ir-exp-dominant}
\delta_{2n,m}=(n-m)\kappa+(1-n)\left(\frac{d}{2}-2\right) \, .
\end{equation}
This relation verifies a posteriori the assumption discussed above
fig.~\ref{fig:gh-DSE-again}: the ghost-gluon scattering kernel is exactly as IR
divergent as to make the ghost-gluon vertex finite in the IR, and this is
independent of the value of the dimension $d$. \\ It remains to discuss higher
order corrections in the skeleton expansion which can be generated via
insertions into given diagrams. The change of the IR exponent due to an exemplary
insertion is given by:
\begin{center}
\begin{minipage}{2.5cm}
\includegraphics[]{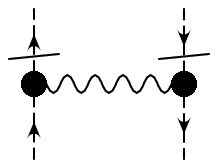}
\end{minipage}
\begin{minipage}{9cm}
$\; (2\kappa+1-\frac{d}{2})+2(-\kappa-1)+2\, \frac{1}{2}+\frac{d}{2}=0,$
\end{minipage}
\end{center}
\vspace*{-4mm}
{\it i.e.\/} the IR-scaling is unchanged. This can also be shown for all other
possible extensions \cite{Huber:2007kc}. Correspondingly, the skeleton
expansion works independent of the number of spacetime dimensions $d$
\cite{vonSmekal:1997is,Zwanziger:2001kw,Lerche:2002ep,Alkofer:2004it,Fischer:2006vf}.
The same holds for the IR limit of the momentum dependent couplings
defined through the different vertices: they only feature a scaling via their
canonical dimension. \\ Finally, let us compare the IR behavior of the
Yang-Mills Greens functions in  different dimensions. We show the IR exponents
of the ghost and gluon propagators and the three- and four-gluon vertices in
table \ref{tb:dim-dep}.
\begin{table}[h]
\centering
\begin{tabular}{l|c|c|c|c}
Dimension &   & {$4$} & {$3$} & {$2$}\\ \hline
$\kappa$ & 			& $0.5953... \approx 0.6$  &  
$0.3976... \approx 0.4$  & $0.2$   \\ \hline
Ghost & $-\kappa-1$ 		& $-1.6$    &  $-1.4$    & $-1.2$  \\ 
Gluon & $2\kappa+1-d/2$ 	        & $\;\;\,0.2$     &  $\;\;\,0.3$     
& $\;\;\,0.4$   \\ 
3-gluon & $-3\kappa +d/2-3/2$      & $-1.3$    &  $-1.2$    & $-1.1$  \\ 
4-gluon & $-4\kappa +d/2 -2$ 	& $-2.4$    &  $-2.1$    & $-1.8$
\end{tabular}
\caption{\label{tb:dim-dep} The dimension dependence of the IR behavior
 of Yang-Mills Green functions. For each $d$ only the continuum solution for 
 $\kappa$ is displayed that has been qualitatively confirmed by lattice 
 simulations.}
\end{table}\\
Whereas the situation in four dimensions is not fully conclusive, yet, the
other solutions have been found in lattice studies in two and three dimensions.
These results for the IR exponents of the propagators, the ghost-gluon vertex
\cite{Cucchieri:2004sq} and the three-gluon vertex in two dimensions
\cite{Maas:2007uv} agree within errors with the corresponding values given in
table \ref{tb:dim-dep}. The qualitative behavior does not
change in different dimensions and even the quantitative values are very
similar. Apart from the listed values of $\kappa$ there is a second branch of
solutions of the DSE equation that has not been seen in lattice simulations. We
will give some additional arguments below that this branch may be unphysical.

\begin{figure}[ht]
\centerline{\includegraphics[width=\textwidth]{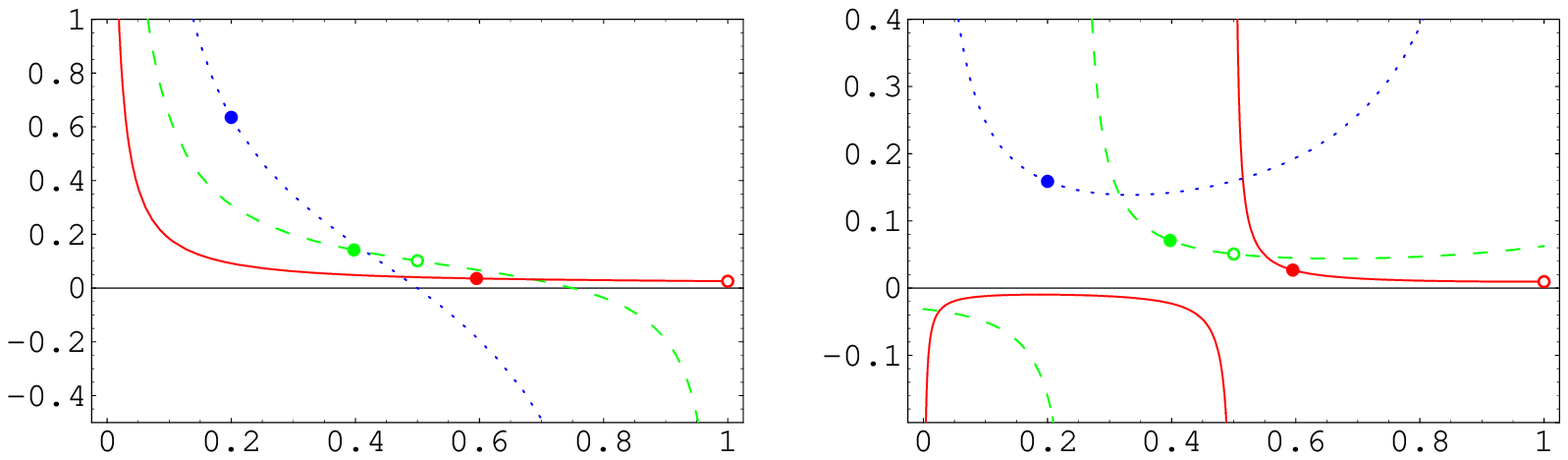}}
\flushleft \vspace*{-2.7cm} $c_{2,0}$ \hspace*{6.9cm} $c_{0,2}$
 \flushleft \vspace*{1.45cm} 
\hspace*{4.0cm} $\kappa$ \hspace*{7.1cm} $\kappa$
\caption{\label{fig:prop-kappa-dep}The $\kappa$-dependence of the
 unrenormalized dressing integrals appearing in the DSEs for the ghost
  (left) and gluon propagator (right). The dotted, dashed and solid lines 
  show the curves for $d=2,3$ and $4$ respectively, see also ref.\ 
  \cite{Lerche:2002ep}. The full and open dots represent the solutions of the 
  first and second branch.}
\end{figure}

\begin{figure}[ht]
\begin{minipage}{0.6\textwidth}
\centerline{\includegraphics[width=0.7\textwidth]{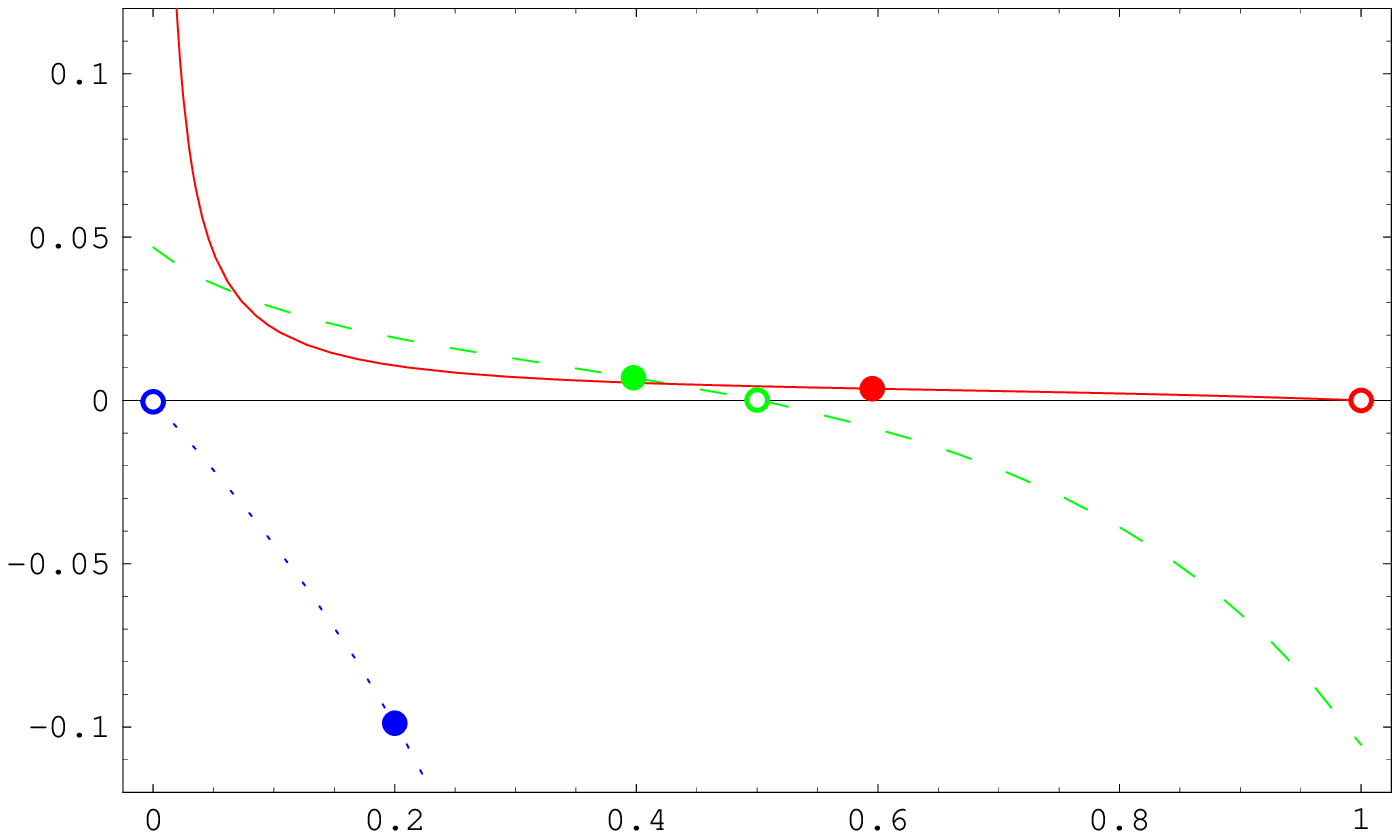}}
\flushleft \vspace*{-3.2cm} \hspace*{0.1cm} $c_{0,3}$ \flushleft \vspace*{2.1cm} 
\hspace*{4.7cm} $\kappa$
\end{minipage}
\begin{minipage}{0.35\textwidth}
\caption{\label{fig:kappa-dep}The $\kappa$-dependence of the overlap of the 
unrenormalized ghost triangle with the tree-level tensor at the symmetric point.
 The dotted, dashed and solid lines show the curves for $d=2,3$ and $4$ 
 respectively. The full and open dots represent the solutions of the first and
  second branch.}
\end{minipage}
\end{figure}

\paragraph*{Dependence of the loop integrals on the IR exponent:}
The DSEs involve loop integrals over bare and dressed vertex functions.  As
discussed the dressing functions exhibit a power law with  appropriate IR
exponents. The corresponding coefficients depend on the IR exponent $\kappa$. 
For the respective coefficients of the propagators, eq.
(\ref{eq:power-laws-props}), which can be computed from the 2-point integral,
eq. (\ref{eq:2-point-integral}), this $\kappa$-dependence is shown in fig.
\ref{fig:prop-kappa-dep}, where $\kappa$-independent prefactors have been
dropped. The interesting point is, that although the curves in fig.
\ref{fig:prop-kappa-dep} differ  considerable, the physical values obtained
from the Dyson-Schwinger solution are far away  from the poles and thereby the
$\kappa$-dependence in their vicinity is rather mild and  qualitatively similar
in each case. 
 \\
In general the vertex integrals feature more  complicated tensor structures
with an increasing number of independent tensor components.  However, there are
general methods to decompose such tensor integrals to standard scalar 
integrals \cite{Davydychev:1991va}. The tensor integrals in the DSE for the
ghost-gluon  and three-gluon vertex reduce to 3-point integrals of the form
\begin{equation}
\label{3int}
  {I}_3 (p,q) \equiv \int \frac{d^d k}{(2\pi)^d} \frac{1}{((k+p)^2)^{\nu_1}} 
  \frac{1}{((k-q)^2)^{\nu_2}} \frac{1}{(k^2)^{\nu_3}}
\end{equation}
A general expression for such scalar one-loop 3-point integrals in arbitrary
dimensions  and with arbitrary powers of the propagators has been obtained in
\cite{Anastasiou:1999ui,Boos:1990rg}. An analytic expression can be obtained
via appropriate series representations for the resulting generalized
hypergeometric functions \cite{IR-vertices,Exton}. Here we will discuss the
dependence of the overlap of the tree-level tensor with the IR-dominant ghost
triangle which presents the IR-leading contribution to the Dyson Schwinger
equation for the three-gluon vertex in  fig. \ref{fig:3g-DSE-skelexp-2}. Since
this loop involves only ghost-gluon vertices that  remain bare to leading
order, its IR behavior can be analyzed semi-perturbatively using  the scaling
form of the dressed ghost propagators, see eq. (\ref{eq:power-laws-props}). We
consider the special kinematic configuration given by the symmetric point
$p^2=q^2=r^2$.  At this point the dressing function of the leading ghost loop
correction to the three-gluon integral scales according to  eq. (\ref{eq:3g})
as $Z_{0,3} = c_{0,3}(\kappa) \cdot (p^2)^{-3\kappa+d/2-2}$ where  the
dependence of the coefficient on the scaling parameter $\kappa$ has been made
explicit.  The coefficient is shown as a function of $\kappa$ in fig.
\ref{fig:kappa-dep}, where the full and open points represent the values for
the first and second branch and $\kappa$-independent factors have been dropped
again.
 \\
The vertex integral has zeros at $\kappa=0$ in two, $\kappa=0.5$ in three and
$\kappa=1$ in four dimensions which coincide precisely with the second branch
of solutions of the DSEs that is not seen in lattice simulations.  Such zeros
in the vertex dressing function would lead to an IR-vanishing coupling. Since
Yang-Mills theory is apparently a strongly interacting theory, this indicates
that these solutions might not be physically relevant. As in the case of the
propagators the $\kappa$-dependence in the vicinity of the solutions of the
first branch is similar in all dimensions.  ynamics would dominate even more. 

\paragraph*{Conclusions:}

As in the 4-dimensional case, the IR behavior of Greens
functions can be extracted via a skeleton expansion in arbitrary dimensions. 
The IR limit of Greens functions is surprisingly insensitive on the
spacetime dimension. As a consequence Yang-Mills theory in lower dimensions
should be qualitatively similar to the 4-dimensional theory, and the
confinement mechanism might possess identical features. Therefore,
corresponding lattice simulations should provide interesting qualitative 
information for the physical case. Such studies in lower dimensions could in
particular provide important lattice results on the IR behavior of vertex
functions. This is particularly challenging due to the observation of
additional kinematic singularities in current DSE studies \cite{IR-vertices}.
\\ 
The results  on the mild $\kappa$-dependence of the remaining DSE solutions
suggest that ghost dominance  is a rather robust mechanism and should not
depend on the details of the employed truncation  scheme. This is further
substantiated by the fact that the main non-linearities in the DSE  system,
which enable the non-trivial fixpoint, arise in the propagator equations
whereas there  is no non-linear feedback to infrared leading order in the vertex
equations. \\

\paragraph*{Acknowledgements:}
We are grateful to A. Cucchieri, A. Maas, T. Mendez and J. Pawlowski for 
interesting discussions. This work has been supported in part by the DFG 
under contract Al279/5-1, by the FWF under contract M979-N16 and by the
 Helmholtz Association 
grant VH-NG-332.


\begin{thebibliography}{99}
\addtolength{\itemsep}{-6pt}
\bibitem{Sternbeck:2005tk}
  A.~Sternbeck {\it et al.}, 
  Phys.\ Rev.\  D {\bf 72} (2005) 014507
  [arXiv:hep-lat/0506007];
  A.~Cucchieri and T.~Mendes,
  Phys.\ Rev.\  D {\bf 73} (2006) 071502
  [arXiv:hep-lat/0602012];
  P.~O.~Bowman {\it et al.},
  arXiv:hep-lat/0703022;
  O.~Oliveira and P.~J.~Silva,
  arXiv:0705.0964 [hep-lat].

\bibitem{Kugo:1979gm}
  T.~Kugo and I.~Ojima,
  Prog.\ Theor.\ Phys.\ Suppl.\  {\bf 66} (1979) 1.

\bibitem{Gribov:1977wm}
  V.~N.~Gribov,
  Nucl.\ Phys.\  B {\bf 139} (1978) 1;
  D.~Zwanziger,
  Nucl.\ Phys.\  B {\bf 364} (1991) 127.

\bibitem{vonSmekal:1997is}
  L.~von Smekal, R.~Alkofer and A.~Hauck,
  Phys.\ Rev.\ Lett.\  {\bf 79} (1997) 3591
  [arXiv:hep-ph/9705242];
C.~S.~Fischer  and R.~Alkofer,
Phys. Lett. {\bf B536}  (2002) 177
[arXiv:hep-ph/0202202].


\bibitem{Zwanziger:2001kw}
  D.~Zwanziger,
  Phys.\ Rev.\  D {\bf 65} (2002) 094039
  [arXiv:hep-th/0109224].

\bibitem{Lerche:2002ep}
  C.~Lerche and L.~von Smekal,
  Phys.\ Rev.\  D {\bf 65} (2002) 125006
  [arXiv:hep-ph/0202194].

\bibitem{Alkofer:2000wg} R.~Alkofer and L.~von Smekal, Phys.\ Rept.\  
{\bf 353} (2001) 281[arXiv:hep-ph/0007355]; \newline
C.~S.~Fischer, J.\ Phys.\ G {\bf 32} (2006) R253 [arXiv:hep-ph/0605173].

\bibitem{Pawlowski:2003hq}
  J.~M.~Pawlowski {\it et al.},
  Phys.\ Rev.\ Lett.\  {\bf 93} (2004) 152002
  [arXiv:hep-th/0312324].

\bibitem{Alkofer:2004it}R.~Alkofer, C.S.~Fischer and F.J.~Llanes-Estrada,
Phys.\ Lett.\  B{\bf 611} (2005) 279[arXiv:hep-th/0412330].

\bibitem{Zwanziger:2003cf}
  D.~Zwanziger,
  Phys.\ Rev.\  D {\bf 69} (2004) 016002
  [arXiv:hep-ph/0303028].

\bibitem{Fischer:2006vf}C.~S.~Fischer and J.~M.~Pawlowski,
Phys.\ Rev.\  D {\bf 75} (2007) 025012
[arXiv:hep-th/0609009].

\bibitem{Bowman:2007du}P.~O.~Bowman {\it et al.}, arXiv:hep-lat/0703022; 
I.~L.~Bogolubsky {\it et al.}, 
arXiv:0707.3611 [hep-lat]; 
A.~Cucchieri, and T.~Mendes, arXiv:0710.0412 [hep-lat].

\bibitem{Fischer:2005ui}
  C.~S.~Fischer, B.~Gruter and R.~Alkofer,
  Annals Phys.\  {\bf 321} (2006) 1918
  [arXiv:hep-ph/0506053];
  C.~S.~Fischer {\it et al.}, Annals Phys. in print,
arXiv:hep-ph/0701050.

\bibitem{Maas:2007uv}
  A.~Maas,
  Phys.\ Rev.\  D {\bf 75} (2007) 116004
  [arXiv:0704.0722 [hep-lat]].

\bibitem{Cucchieri:2006tf}A.~Cucchieri, A.~Maas and T.~Mendes, 
Phys.\ Rev.\ D {\bf 74} (2006) 014503
[arXiv:hep-lat/0605011];
  A.~Cucchieri, T.~Mendes and A.~R.~Taurines,
  Phys.\ Rev.\  D {\bf 67} (2003) 091502
  [arXiv:hep-lat/0302022].

\bibitem{Taylor:1971ff}J.~C.~Taylor, Nucl.\ Phys.\  B {\bf 33} (1971) 436.

\bibitem{Anastasiou:1999ui}C.~Anastasiou, E.~W.~N.~Glover and C.~Oleari, 
Nucl.\ Phys.\ B {\bf 572} (2000) 307
[arXiv:hep-ph/9907494].

\bibitem{Huber:2007kc}
  M.~Huber {\it et al.}, 
  arXiv:0705.3809 [hep-ph].

\bibitem{Cucchieri:2004sq}
  A.~Cucchieri, T.~Mendes and A.~Mihara,
  JHEP {\bf 0412} (2004) 012
  [arXiv:hep-lat/0408034].


\bibitem{Davydychev:1991va}
  A.~I.~Davydychev,
  Phys.\ Lett.\  B {\bf 263} (1991) 107.

\bibitem{Boos:1990rg}E.~E.~Boos and A.~I.~Davydychev, 
Theor.\ Math.\ Phys.\  {\bf 89} (1991) 1052.

\bibitem{IR-vertices} R.~Alkofer, M.~Huber and K.~Schwenzer, in preparation.

\bibitem{Exton}
H. Exton, J. Phys. A: Math. Gen. 28 (1995) 631.


\end{thebibliography}
\end{document}